\documentclass[%
 aip,
 jmp,%
 amsmath,amssymb,
 reprint,%
]{revtex4-1}
\usepackage{graphicx}% Include figure files
\usepackage{dcolumn}% Align table columns on decimal point
\usepackage{bm}% bold math
\usepackage{epstopdf}
\begin{document}
\preprint{APS/123-QED}
\title{Deflection of light ray due to a charged body using Material Medium Approach}% Force line breaks with \\
\author{Saswati Roy}
 \email{sr.phy2011@yahoo.com}
 \affiliation{%
 Physics Department, Assam University, Silchar-788011\\
 India }%
 \author{A.K.Sen}%
\email{asokesen@yahoo.com}
\affiliation{%
 Physics Department, Assam University, Silchar-788011\\
 India }%
\date{\today}% It is always \today, today,
\begin{abstract}
 The gravitational deflection of light ray is an important
 prediction of General Theory of Relativity. In this paper
 we develop an analytical expression of the deflection of light
 ray without any weak field approximation due to a charged
 gravitating body represented by Reissner-Nordstr\"{o}m (RN)
 and Janis-Newman-Winicour (JNW) space time geometry,
 using material medium approach. It is
 concluded that although both the geometries represent the
 charged, non-rotating, spherically symmetric gravitating body,
 but the effect of charge on the gravitational deflection
 is just opposite to each other. The gravitational deflection decreases with
 charge in the RN geometry and increases with charge in the
 JNW geometry. The calculations obtained here are compared with other
 methods done by different authors.
 The formalism is applied to an arbitrary selected
 pulsar $PSR B 1937+21$ as a gravitating body, as a test case.

\begin{description}
\item[PACS numbers]
95.30.Sf, 04.70.-s, 04.20.Jb, 04.25.D
\end{description}
\end{abstract}
%\pacs{95.30.Sf, 04.70.-s, 04.20.Jb, 04.25.D}% PACS, the Physics and Astronomy
%Use showkeys class option if keyword
\maketitle
%\tableofcontents
%\end{center}
\section{\label{1}Introduction}	

 One of the striking predictions of General Theory of Relativity
 is the deflection of light ray in the presence of a gravitating mass.
 The predictions of gravitational effect on light, began by Einstein in 1913
 and was confirmed by Eddington in 1919, during the total solar eclipse.

In this paper we have used the
\emph{Material Medium Approach}
to find the deflection of light ray
  due to a charged
 gravitating body represented by Reissner-Nordstr\"{o}m (RN)
 space-time and Janis-Newman-Winicour (JNW) space time,
 without assuming any weak field approximation. The material medium approach
 is different from conventional null geodesic approach, through which
 the deflection of light ray can be calculated. The material medium approach has been used
 by several researchers (details in Section \ref{2}), where the effect of gravity
 is viewed as a change in the refractive index of the medium through which light is travelling.

The Reissner-Nordstr\"{o}m (RN) metric\cite{jr1,jr2} and
Janis-Newman-Winicour (JNW) metric\cite{jr3} both are static solution of
Einstein-Maxwell field equation for a charged, non-rotating,
spherically symmetric gravitating body. The RN solution has an event horizon and
a Cauchy horizon but JNW solution has a curvature singularity and a naked singularity.
If there is no charge, both the solutions reduce to the Schwarzschild solution.

The concept of scalar field theory came before the general theory of relativity.
In 1956 O. Bergmann\cite{jr4} discussed the scalar field theory as a theory of Gravitations
and in 1957 O. Bergmann and R. Leipnik\cite{jr5} discussed the field equations
of a static spherically symmetric scalar field. Many authors\cite{jr6,jr7,jr8,jr9} studied the
field of charged particles in General Relativity.

Role of scalar field in Gravitational lensing by JNW black hole was discussed by Virbhadra {\it et.al}\cite{jr10}
in 1998. In 2002 Eiroa {\it et.al}\cite{jr11}
discussed about the Gravitational lensing by RN
black hole in strong field limit. The two groups of
authors studied the null geodesics in a general static spherically symmetric spacetime
and calculated the Einstein's deflection angle. Further they discussed about the image formation
and lensing effect. In 2000 Virbhadra and Ellis\cite{jr12} first defined the
photon sphere which is the starting point of the strong field limit expansion and in the next paper
Claudel, Virbhadra and Ellis\cite{jr13} discussed the geometry of photon surfaces
and calculated the radius of the photon sphere in different space-time. In 2002,
Virbhadra and Ellis\cite{jr14} discussed the gravitational lensing by naked singularities.
In the same year, Bozza\cite{jr15} extended the analytical theory of strong
lensing for  a general class of static spherically symmetric matrices.
In 2003 Bhadra\cite{jr16} also discussed the lensing effect in the strong gravity
regime due to a charged black hole in case of strong field limit.
Recently, Amore and Arceo in 2006\cite{jr17} considered the gravitational
lensing for different black holes including RN metric type and JNW metric type.
In 2012 Chowdhury \textit{et.al}\cite{jr18} studied the circular geodesics in the
JNW and Gamma metric space-time. In 2014 Chakraborty and SenGupta\cite{jr19}
estimated the perihelion precession and bending of light due
to charged black hole using RN metric and showed that
perihelion shift decreases with the increase of charge
and bending of light is almost similar to that by Schwarzschild
field at a large distance from the source.

The present paper is organized as follows: In section \ref{2.1} we describe the
RN and JNW space time.  In section \ref{2.2} we obtain
the velocity and refractive index due to RN and JNW geometry and explain the effects by plotting
graphs.
In a separate section \ref{2.3} we obtain the deflection angle and describe the results obtained by RN
and JNW geometry. In section \ref{2.4} we
compared the refractive index and bending angle due to a charged
gravitating body for both the space time with other methods and by other authors. Finally in section \ref{3}
we make discussion and draw conclusions on our result.

\section{{\label{2}}Optical medium approach}

The deflection of light ray can be obtained by the most general method
as null geodesic. As already outlined in section \ref{1},
the Optical Medium Approach is an alternate method by
which also we can calculate the gravitational deflection of light ray.
In this method the gravitational effect is represented by an equivalent refractive
index of the medium. Thus if a ray of light passes through a material
medium, the light ray will deviate due to the variation of the refractive index of
the corresponding media. This method was first introduced by Tamm\cite{jr20} in 1924.
Balaz\cite{jr21} in 1958 considered that if an electro magnetic wave passed through
the gravitational field of an rotating body, then the polarization vector gets rotated.
In 1960 Plebanski\cite{jr22} studied the scattering
of light ray by gravitational field. On the level of geometrical optics
the author formulated the generalized form of the Einstein's deflection angle
and examined the direction of the plane of polarization.
Felice\cite{jr23} in 1971 used the concept of equivalent material medium to
deduce the refractive index and the deflection angle of a light ray in a
static and spherically symmetric space time. B. Mashhoon\cite{jr24,jr25}
obtained the scattering cross section and polarization of the scattered
wave by Schwarzschild and Kerr gravitating body.
Fischbach and Freeman\cite{jr26} in 1980
used the same approach and obtained the refractive index and deflection upto 2nd order
in Schwarzschild geometry. Evans, Rosenquist, Nandi and Islam\cite{jr27,jr28,jr29,jr30}
used the Fermat's principle to calculate the
effective refractive index to derive the deflection angle in Schwarzschild geometry
for massless and massive particles.
P.M. Alsing\cite{jr31} in 1998 extended the formalism of Evans, Nandi and Islam\cite{jr30}
in case of Kerr field geometry. Sereno\cite{jr32,jr33} is also used the Fermat's principle to
discuss the gravitational lensing and Faraday rotation in the weak field limit.
In 2004, Sereno\cite{jr34} discussed
the gravitational lensing in weak field limit of RN metric and
used Fermat's principle to calculate the deflection angle.
Ye and Lin\cite{jr35} in 2008 discussed the exterior and interior solution
of the refractive index for a static spherically symmetric gravitational field and
found the most general formula of refractive index in terms of potential
in weak field limit and also discussed the effect of lensing.

Very recently \emph{Material Medium Approach} was used by Sen\cite{jr36} and Roy and Sen\cite{jr37}
to calculate the light deflection angle due to a gravitating body
represented by Schwarzschild and Kerr space-time geometry respectively.

\subsection{{\label{2.1}}Reissner-Nordstr\"{o}m (RN) space-time and
Janis-Newman-Winicour (JNW) space-time}

The Reissner-Nordstr\"{o}m space-time is the
spherically symmetric solution of coupled equations of Einstein
and of Maxwell. A non-rotating black hole with gravitational mass $m$
and a charge length $Q$ can be represented by a Reissner-Nordstr\"{o}m(RN)\cite{jr1,jr2}
line element. In Boyer-Lindquist co-ordinates $(ct,r,\theta,\phi)$\cite{jr38}
the RN metric is in the form\cite{bk1}:
\begin{widetext}
\begin{eqnarray}\label{E1}
 ds^{2}=&&  (1-\frac{2 m}{r} + \frac{Q^2}{ r^2})c^2 dt^2 - (1-\frac{2 m}{r} + \frac{Q^2}{r^2})^{-1} dr^2 - r^2 (d\theta^2 +\sin^2\theta d\phi^2)
\end{eqnarray}
\end{widetext}

where $Q^2=\frac{G e^2}{4 \pi \epsilon_0 c^4}$  and
                                                    $e$ is the scalar charge,
                                                    $\frac{1}{4 \pi \epsilon_0}$ is the coulomb force-constant,
                                                    $G$ is the gravitational constant,
                                                    $c$ is the speed of light.
$Q$ has the dimension of length. Further, $2m = r_g$ and $r_g$ is the Schwarzschild radius = $\frac{2GM}{c^2}$.

This metric has an event horizon at $r_+ = m + \sqrt{m^2-Q^2}$  and
a Cauchy horizon at $r_- = m - \sqrt{m^2-Q^2}$. For $Q^2 > m^2$ , $r_+$
or $r_-$ has no real solution and hence $Q$ has a limit as $Q^2 \leq m^2$.
Thus event horizon exists for  $0 \leq Q^2 \leq m^2$
and Cauchy horizon exists for $0 < Q^2 \leq m^2$.

Photon sphere is a region where light travels in close orbits due to
strong gravitational effect of the gravitating body. Thus
it is the minimum radius of the stable orbit.
The radius of the photon sphere of this metric is\cite{jr13}

\begin{equation}\label{E2}
r_{ps}^\pm = \frac{3m\pm \sqrt{9 m^2- 8 Q^2}}{2}
\end{equation}

Thus the surface $S^+$ for radius $r_{ps}^+$ exists for
$ 0 \leq Q^2 \leq 9/8 m^2 $.
And the surface $S^-$ for radius $r_{ps}^-$ exists for
$ 0 < Q^2 \leq 9/8 m^2 $. Both the surfaces $S^+$ and $S^-$
coincide for $Q^2 = 9/8 m^2 $.

Jenis-Newman-Winicour (JNW) space-time\cite{jr3} on the other hand also represents
the most general spherically symmetric, static and asymptotically flat
solution of Einstein's field equation which is coupled to a massless
scalar field. The JNW solution in the co-ordinates ($ct$, $r$, $\theta$, $\phi$)
can be represented by the line element\cite{jr10}
\begin{widetext}
\begin{equation}\label{E3}
ds^2=(1-\frac{l}{r})^\gamma c^2 dt^2 - (1-\frac{l}{r})^{-\gamma} dr^2 - r^2 (1-\frac{l}{r})^{1-\gamma} (d\theta^2 + sin^2\theta d\phi^2)
\end{equation}
\end{widetext}
with scalar field

\begin{equation}\label{E4}
\Phi = \frac{Q}{l\sqrt{4 \pi}} \ln(1-\frac{l}{r})
\end{equation}

where

\begin{subequations}
\begin{equation}\label{E5a}
l = 2 \sqrt{m^2 + Q^2}
\end{equation}
and
\begin{equation}\label{E5b}
\gamma = \frac{2m}{l} = \frac{m}{\sqrt{m^2 + Q^2}}
\end{equation}
\end{subequations}

When $Q=0$, $ \gamma=1 $ and finally the JNW metric also goes to the
Schwarzschild metric. The JNW metric has a curvature singularity
at $r=l$ i.e. $r$ has a limit as $l < r < \infty$.

The radius of the photon sphere of this geometry is\cite{jr13}

\begin{eqnarray}\label{E6}
r_{ps} =&& \frac{l(1 + 2\gamma)}{2}\nonumber\\
=&&  \sqrt{m^2 + Q^2} (1 + \frac{2m}{ \sqrt{m^2 + Q^2}})\nonumber\\
=&&  2m + \sqrt{m^2 + Q^2}
\end{eqnarray}

which exists only for $\frac{1}{2} < \gamma \leq 1 $
i.e. for $0 \leq Q^2 < 3m^2$ which is mentioned as a
weak naked singularity. If we consider the value of
$\gamma \leq 1/2$ then we will get the strong naked singularity.
Naked singularity may or may not be within photon sphere. If the singularity point
is within photon sphere then it is called as weak naked
singularity. If the singularity point is not covered by
any photon sphere then it is called as strong naked singularity\cite{jr13}.

 \subsection{\label{2.2}Refractive index as calculated in RN and JNW space-time}

By following the same procedure as Sen\cite{jr36} and Roy and Sen\cite{jr37} we can get the
isotropic form of the line element in the field of RN  space-time and
JNW space-time in terms of ($ct$, $\rho$, $\theta$, $\phi$) as
\begin{widetext}
\begin{equation}\label{E7}
ds^2=   \frac{ (1-\frac{m^2-Q^2}{4 \rho^2})^2}{(1+\frac{m}{ \rho}+\frac{m^2-Q^2}{4\rho^2})^2}c^2 dt^2 -
(1+\frac{m}{ \rho}+\frac{m^2-Q^2}{4\rho^2})^2  \{d\rho^2 + \rho^2  (d\theta^2 + \sin^2\theta d\phi^2)\}
\end{equation}
and
\begin{equation}\label{E8}
ds^2 = (\frac{1-l/4\rho}{1+l/4\rho})^{2\gamma} c^2 dt^2 - (1+\frac{l}{4\rho})^{2(1+\gamma)} (1-\frac{l}{4\rho})^{2(1-\gamma)} (d\rho^2 + \rho^2 (d\theta^2 + sin^2\theta d\phi^2))
\end{equation}
\end{widetext}
respectively.

To get the isotropic form of the line element in terms of
($ct$, $\rho$, $\theta$, $\phi$) we have introduced a new co-ordinate as

\begin{subequations}
\begin{equation}\label{E9a}
\rho=\frac{1}{2}(r- m \pm \sqrt{r^2-2 m r +Q^2} )
\end{equation}
or
\begin{equation}\label{E9b}
r=\rho(1+\frac{m}{ \rho}+\frac{m^2-Q^2}{4\rho^2})
\end{equation}
\end{subequations}

for RN space-time and

\begin{subequations}
\begin{equation}\label{E10a}
\rho=\frac{1}{2} [(r-\frac{l}{2}) + r^{\frac{1}{2}} (r-l)^{\frac{1}{2}}]
\end{equation}
or
\begin{equation}\label{E10b}
r=\rho (1+\frac{l}{4\rho})^{2}
\end{equation}
\end{subequations}

for JNW space-time.

 In a spherical coordinate system $ds^2 = f (\rho)dt ^2 - d\overrightarrow{ \rho} ^2 $
 where the quantity $d\overrightarrow{\rho} ^2=\{d\rho^2 + \rho^2  (d\theta^2 + \sin^2\theta d\phi^2)\}$
has the dimension of the square of the infinitesimal length vector $\overrightarrow{\rho}$.
 Thus, by setting $ds=0$, the velocity of light
can be identified  as $v(\rho)= \sqrt {f(\rho )}$ (Sen\cite{jr36}).
Now from the above isotropic expression the velocity of light ray in terms of $\rho$ can be written as:

\begin{eqnarray}\label{E11}
v(\rho)=\frac{ (1-\frac{m^2-Q^2}{4 \rho^2})}{(1+\frac{m}{ \rho}+\frac{m^2-Q^2}{4\rho^2})^2 }c
\end{eqnarray}

for RN space-time.

Again

\begin{equation}\label{E12}
    v(\rho) = \frac{(1 - \frac{l}{4\rho}) ^{2\gamma -1}}{(1 +\frac{l}{4\rho}) ^{2\gamma + 1}} c
\end{equation}

for JNW space-time.

So with the value of $\rho$ as in Eqn.
(\ref{E9a})  and (\ref{E10a}), the velocity $v(r,Q)$ in terms of $r$ and $Q$ becomes

\begin{eqnarray}\label{E13}
v(r,Q)=&& \frac{r^2-r r_g +Q^2}{r^2}c\nonumber\\
=&& (1- \frac{r_g}{r} + \frac{Q^2}{r^2})c\nonumber\\
=&& (1 - \frac{r_g}{r})c + \frac{Q^2}{r^2}c
\end{eqnarray}

The first term of the above expression refers to the
velocity of the light ray due to Schwarzschild geometry \cite{jr36}
and the second term is due to the charge under RN geometry.

and

\begin{eqnarray}\label{E14}
v(r,Q) &=& (\frac{r-l}{r})^\gamma c\nonumber\\
&=& (\frac{r - \sqrt{r_g^2 + 4 Q^2}}{r})^ {\frac{r_g}{\sqrt{r_g^2 + 4 Q^2}}}c \nonumber\\
&=& ( 1 - \frac{r_g}{r}\sqrt{1 + 4 \frac{Q^2}{r_g^2}} )^ {\frac{1}{\sqrt{1 + 4 \frac{Q^2}{r_g^2}}}}c
\end{eqnarray}

where we have substituted the value of $l=2\sqrt{m^2+Q^2}$, $\gamma = \frac{2m}{l}$ and $m=\frac{r_g}{2}$.

Therefore the refractive index of light ray, $n(r,Q)$ can be expressed by the relation

\begin{eqnarray}\label{E15}
n(r,Q)=\frac{ 1 }{(1 - \frac{r_g}{r}) + \frac{Q^2}{r^2}}
\end{eqnarray}

for RN space-time.

And

\begin{eqnarray}\label{E16}
    n(r,Q)    = (\frac{1}{ 1 - \frac{r_g}{r}\sqrt{1 + 4 \frac{Q^2}{r_g^2}}} )^ {\frac{1}{\sqrt{1 + 4 \frac{Q^2}{r_g^2}}}}
\end{eqnarray}

for JNW space-time.

Replacing $r/r_g$ by $x$ and $Q/r_g$ by $q$, the above
expression of refractive indices become:

\begin{eqnarray}\label{E17}
n(x,q)=&& \frac{1}{1-\frac{1}{x}+\frac{q^2}{x^2}}\nonumber\\
=&& \frac{1}{1-\frac{1}{x}} [1+ \frac{q^2}{x^2(1-\frac{1}{x})}]^{-1}\nonumber\\
=&& \frac{x }{x-1}  [1+ \frac{q^2}{x(x-1)}]^{-1}\nonumber\\
=&&  n_{0}(x)  [1+  C_{x}]^{-1}
\end{eqnarray}

for RN space-time. In the above, we introduced the parameters
$n_{0}(x)=\frac{x}{x-1}$ and $C_{x} =  \frac{q^2}{x(x-1)}$.
We can show for all $r >> r_g$ and $r >> Q $ , we must
 have $x >> 1$ and $x >>q$. Now as $x >> 1$, we can approximate $(x-1)\sim x$ and then
 we can finally show that $C_x << 1$.

And

\begin{eqnarray}\label{E18}
n(x,q)= (\frac{1}{1- \frac{1}{x} \sqrt{1+4q^2}})^{\frac{1}{\sqrt{1+4q^2}}}
\end{eqnarray}

for JNW space-time.

Now at $Q=0$ (or $q=0$), the Reissner-Nordstr\"{o}m  space-time and
Janis-Newman-Winicour space-time, both become the
Schwarzschild space-time, so that the velocity and
refractive index  become

\begin{eqnarray}\label{E19}
v(x)= (1-\frac{1}{x})c =\frac{r-r_g}{r}c
\end{eqnarray}
and
\begin{eqnarray}\label{E20}
n(x)= \frac{x}{x-1}= \frac{r}{r-r_g}
\end{eqnarray}

which are exactly same as the velocity and refractive index
calculated by Sen\cite{jr36} for Schwarzschild space-time.

However, Fischbach and Freeman \cite{jr26} have also calculated the refractive index
in terms of PPN parameters where the photon is propagating in a Minkowskian
space-time but with a local index of refraction. But in our process, we converted the
metric into isotropic form (in terms of $\rho$) by co-ordinate transformation
to get a general expression of index of refraction. Fischbach and Freeman \cite{jr26}
defined the index of refraction as an infinite convergent series as

\begin{eqnarray*}
n(r)=1+A/r+B/r^{2}+.....
\end{eqnarray*}

with $A=r_{g}$ and $B=f(r_{g})$
where as the expression for Schwarzschild co-ordinate by Sen \cite{jr36}
can also be represented by the infinite convergent series as

\begin{eqnarray*}
n(r)=1+(r_{g}/r)+(r_{g}/r)^{2}+(r_{g}/r)^{3}+........
\end{eqnarray*}

Thus both the expressions are same in the weak field limit.

Now we will study the variation of refractive index ($n(x,q)$) as a function of $q$.
As a test case, we consider a pulsar \textit{PSRB 1937+21}\cite{jr39}
as the charged gravitating body.
This pulsar is of mass 1.35$M_\odot$, it has time period 1.557 ms and
physical radius 20.2 \textit{Km} (Nunez {\it et al.}\cite{jr40}).

\begin{figure*}[!htb]\centering
\includegraphics{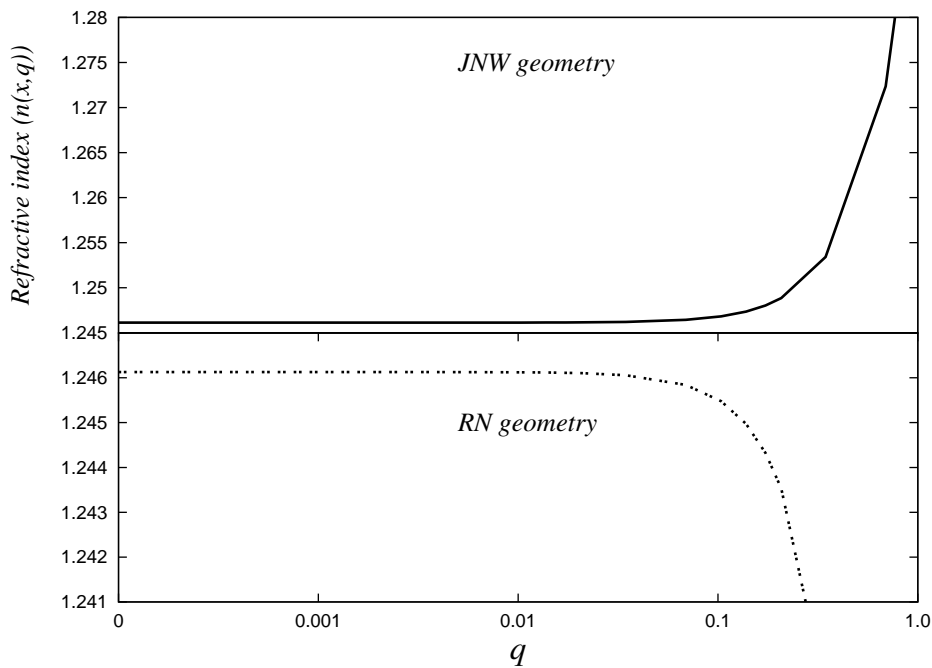}% Here is how to import EPS art 3
\caption{Refractive index($n(x,q)$)  as a function of $q$ for an arbitrarily selected pulsar PSRB 1937+21.}\label{fig:1}
\end{figure*}

Fig. \ref{fig:1} shows the variation of refractive index ($n(x,q)$)
with normalized charge radius $q$ in RN geometry and JNW geometry. RN geometry explains the
decrease of refractive index with the increase of $q$
and at $q=0$ the value of refractive index is maximum.
According to RN geometry, it has two horizons and accordingly has
some limitation to choose the value of $Q$ as $Q^2 \leq m^2 $ or $ Q^2 \leq 0.25 r_g^2$
so that $q^2 \leq 0.25$.
Here we have chosen the value of $Q$ accordingly for the
given pulsar, so that the maximum value is $q=0.5$ or $Q=1.9949$ km as the value
of Schwarzschild radius, $r_g$ of that pulsar is approximately 3.9898 km.

In JNW geometry with the increase of $q$, refractive index increases
and at $q=0$ the value of refractive index is minimum.
JNW metric has a curvature singularity at $r=l$
and a naked singularity at $1/2 < \gamma \leq 1$ which gives
$0 \leq Q^2 < 3m^2$ or $0 \leq Q^2 < 0.75 r_g^2$,
so that $q^2  < 0.75$ (expression for $l$ and $\gamma$ are given
in Eqn.(\ref{E5a}) and (\ref{E5b})). We have also considered the maximum value of $Q$
as $3.4552$ km or $q=0.8660$ remembering such conditions.

\subsection{\label{2.3}Calculation of deflection of light ray from refractive index}

Using the expression of refractive index the trajectory of light
ray can be written as\cite{jr35,jr36,jr37,bk2}:

\begin{equation}\label{E21}
\triangle \psi = 2 \int^{\infty}_{b} { \frac {dr}{r
\sqrt{(\frac{n(r).r}{n(b).b})^2-1}}}- \pi
\end{equation}

As reported earlier by Sen\cite{jr36} and Roy and Sen\cite{jr37}, in the
present paper also we are considering that the light is
approaching from asymptotic infinity ($r=-\infty$) towards the
gravitating body and then it goes to $r=+\infty$ after undergoing
certain amount of deflection ($\triangle \psi$). Here,the
gravitating body is characterized by the Schwarzschild radius $r_g$
and the charge length $Q$. Here the closest distance of approach
or the impact parameter ($b$) is considered as the physical radius
of the gravitating body. When the light ray passes through the
closest distance of approach, the tangent to the trajectory
becomes perpendicular to the vector ${\overrightarrow r}$
(which is  ${\overrightarrow b}$).

Now we change the variable from $r$ to $x=\frac{r}{r_g}$ according as the
Roy and Sen\cite{jr37}, so that $dr=r_g  dx$ and the corresponding limit
changes to $x=\frac{b}{r_g}=v$ and $x=\infty$,
as the limit of $r$ changes from $r=b$ and $r=\infty$.

Therefore, the value of deflection  can be written as :

\begin{eqnarray}\label{E22}
\triangle \psi &=& 2 n(v,q) v \int^{\infty}_{v} { \frac {dx}{x
\sqrt{(n(x,q) x)^2-(n(v,q) v)^2}}}- \pi\nonumber\\
&=& 2 I - \pi
\end{eqnarray}

where

\begin{eqnarray}\label{E23}
I&=&n(v,q) v \int^{\infty}_{v}  { \frac {dx}{x\sqrt{(n(x,q) x)^2-(n(v,q) v)^2}}}\nonumber\\
&=& D \int^{\infty}_{v}  { \frac {dx}{x\sqrt{(n(x,q) x)^2- D^2}}}
\end{eqnarray}

where we have substituted $n(v,q) v = D$.

\subsubsection{\label{2.3.1} RN space time}

The general expression of refractive index due to charged body in Reissner
Nordstr\"{o}m space time is represented by equation(\ref{E15}) or (\ref{E17}). Now,
by substituting the value of $n(x,q)$
in the above equation (\ref{E23})
we get the deflection of light ray in the field of Reissner-Nordstr\"{o}m
space-time as

\begin{widetext}
\begin{eqnarray}\label{E24}
\triangle \psi &=& (\frac{D_{r}}{D_{0}}-1) \pi   + 2D_{r}\{\int ^{a}_{0}\frac{ z dz}{ \sqrt{1-D_{0}^2z^{2}(1-z)^{2}}}\nonumber\\
&&-  \frac{1}{2} \int^{a}_{0} \frac{( 1 -z)}{\sqrt{   1 -D_{0}^2 z^{2}(1 -z)^{2}}}
[\frac{ -   \frac{2q^2z^2 (1-z)+q^4z^4}{((1-z)+q^2z^2)^2}  +D_{0}^2 z^2(1-z)^2\{\frac{2q^2v(v-1)+q^4}{(v(v-1)+q^2)^2}\} }{ 1-D_{0}^2 z^2(1-z)^2}]dz\nonumber\\
&& + \frac{3}{8} \int^{a}_{0} \frac{( 1 -z)}{\sqrt{   1 -D_{0}^2 z^{2}(1 -z)^{2}}}
[\frac{ -   \frac{2q^2z^2 (1-z)+q^4z^4}{((1-z)+q^2z^2)^2}  +D_{0}^2 z^2(1-z)^2\{\frac{2q^2v(v-1)+q^4}{(v(v-1)+q^2)^2}\} }{ 1-D_{0}^2 z^2(1-z)^2}]^2 dz\nonumber\\
&& -  \frac{5}{16} \int^{a}_{0} \frac{( 1 -z)}{\sqrt{   1 -D_{0}^2 z^{2}(1 -z)^{2}}}
[\frac{ -   \frac{2q^2z^2 (1-z)+q^4z^4}{((1-z)+q^2z^2)^2}  +D_{0}^2 z^2(1-z)^2\{\frac{2q^2v(v-1)+q^4}{(v(v-1)+q^2)^2}\} }{ 1-D_{0}^2 z^2(1-z)^2}]^3 dz\nonumber\\
&& +  \frac{35}{64} \int^{a}_{0} \frac{( 1 -z)}{\sqrt{   1 -D_{0}^2 z^{2}(1 -z)^{2}}}
[\frac{ -   \frac{2q^2z^2 (1-z)+q^4z^4}{((1-z)+q^2z^2)^2}  +D_{0}^2 z^2(1-z)^2\{\frac{2q^2v(v-1)+q^4}{(v(v-1)+q^2)^2}\} }{ 1-D_{0}^2 z^2(1-z)^2}]^4 dz\nonumber\\
&& -  \frac{63}{256} \int^{a}_{0} \frac{( 1 -z)}{\sqrt{   1 -D_{0}^2 z^{2}(1 -z)^{2}}}
[\frac{ -   \frac{2q^2z^2 (1-z)+q^4z^4}{((1-z)+q^2z^2)^2}  +D_{0}^2 z^2(1-z)^2\{\frac{2q^2v(v-1)+q^4}{(v(v-1)+q^2)^2}\} }{ 1-D_{0}^2 z^2(1-z)^2}]^5 dz\nonumber\\
&&+ ........\}
\end{eqnarray}
\end{widetext}

To get the above equation (\ref{E24}) from equation (\ref{E23}) we have done some algebraic calculations
and the details of the calculation are given in Appendix \ref{A}.

\subsubsection{\label{2.3.2} JNW space time}

The refractive index in JNW space time is expressed  by the equation(\ref{E16}) or (\ref{E18}).
Thus with the value of refractive index ($n(x,q)$) and equation (\ref{E23})
the value of deflection of light ray in the field of JNW space time as

\begin{equation}\label{E25}
\triangle \psi =  2 D_j \int^{a}_{0} { \frac{ z (1- z \sqrt{1+4q^2})^{\frac{1}{ \sqrt{1+4q^2}}-1} }{ \sqrt{1 -  D_j^2 z^2 (1- z \sqrt{1+4q^2})^{2/\sqrt{1+4q^2}}}}} dz
\end{equation}

Here also the details of the calculation are given in Appendix \ref{B}.

\begin{figure*}[!htb]\centering
\includegraphics{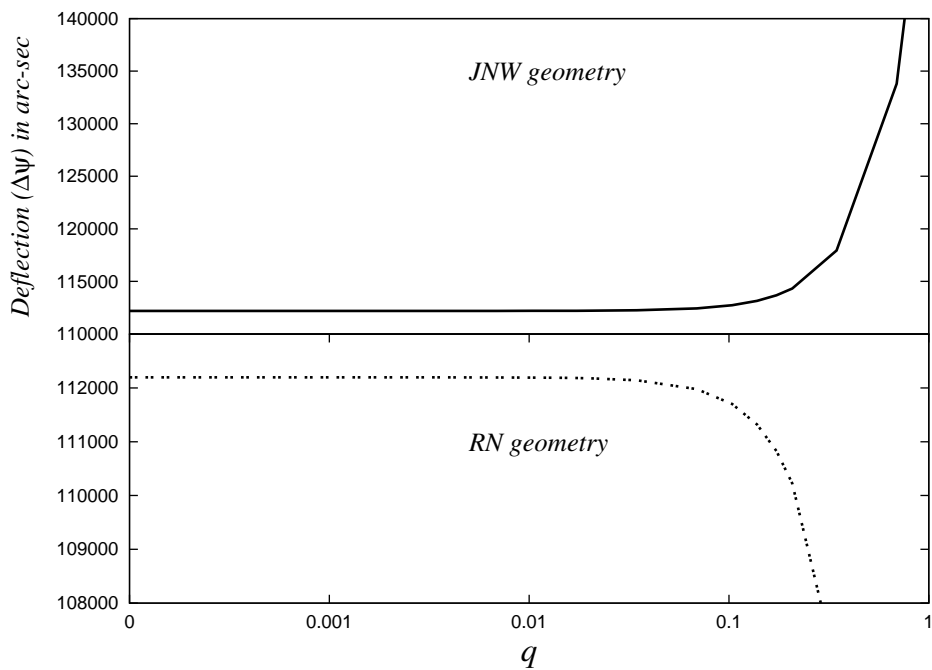}% Here is how to import EPS art 3
\caption{Deflection($\triangle\psi$) as a function of $q$ for an arbitrarily selected pulsar PSRB 1937+21.}\label{fig:2}
\end{figure*}

The above expressions (\ref{E24}) and (\ref{E25}) have been obtained
without applying any weak field approximation at any stage. Thus,
we may consider, these are the exact expressions of deflection of light ray
in RN and JNW space-time
using an equivalent material medium approach.

Here also if we apply the boundary condition as $q=0$, both the expressions
(\ref{E24}) and (\ref{E25}) exactly match with the bending angle
due to Schwarzschild metric as given by Sen\cite{jr36}.

Considering the impact parameter as the physical radius of the gravitating body
the variations of deflection($\triangle\psi$)  as a function of charge radius $q$,
for RN and JNW geometry are shown in Fig. \ref{fig:2}. It also shows that as the value
of the $q$ increases, the value of deflection decreases in RN geometry and increase in JNW geometry.

\begin{figure*}[!htb]\centering
\includegraphics{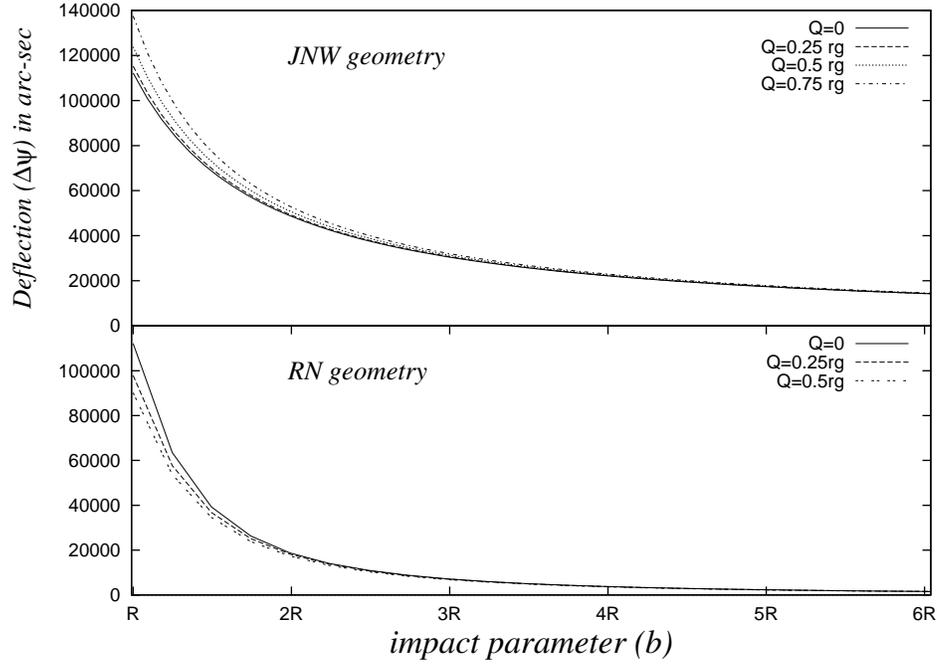}% Here is how to import EPS art 3
\caption{Deflection($\triangle\psi$)  as a function of impact parameter($b$) for an arbitrarily selected pulsar PSRB 1937+21.}\label{fig:3}
\end{figure*}

The variations of deflection($\triangle\psi$)  as a function of impact parameter ($b$)
for different values of $Q$ are shown in Fig. \ref{fig:3}.
Figure shows the variation in RN geometry and JNW geometry
in lower and upper panel respectively. For RN geometry
we have considered the value of $Q$ as $ 0, 0.5 m, 1.0 m$
(as $Q^2 \leq m^2$) or $0, 0.25 r_g, 0.5 r_g$. And for
JNW geometry the value of $Q$ as $ 0, 0.5 m, 1.0 m, 1.5m$
(as $0 \leq Q^2 < 3m^2$) or $ 0, 0.25 r_g, 0.5 r_g, 0.75 r_g$.
As the value of $b$ increases i.e. as we move towards asymptotically flat space,
these curves merge into each other. The deflection with $Q=0$ indicates the
deflection with respect to the Schwarzschild geometry.

\subsection{\label{2.4}Comparison with other recent work}

In our present work, we obtained the refractive index and angle of
deflection of light ray due to RN space-time and JNW space-time
following \textit{Material Medium Approach}. We can compare our
obtained results with others recent work.

\begin{figure*}[!htb]\centering
\includegraphics{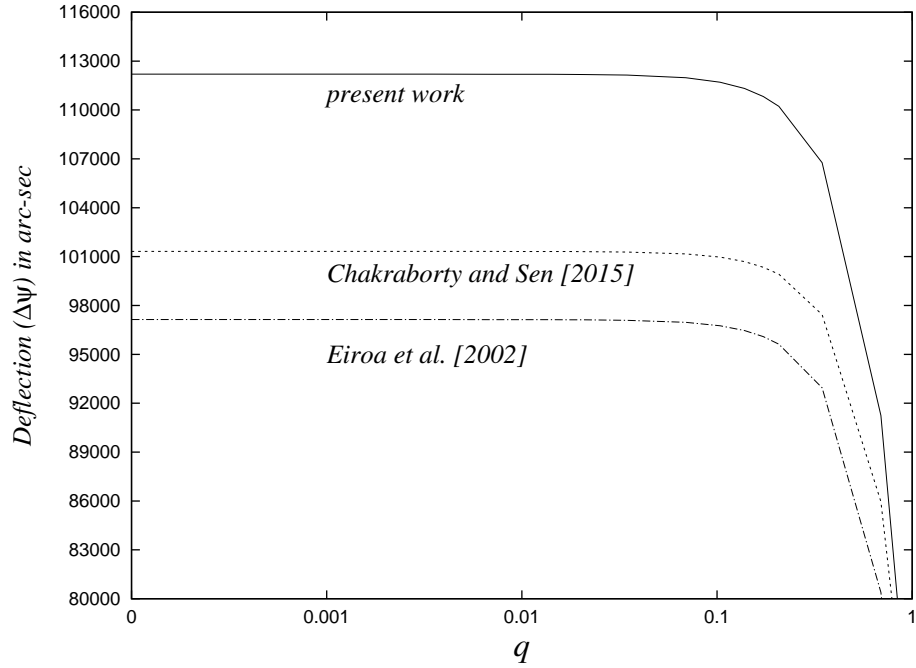}% Here is how to import EPS art 3
\caption{Deflection($\triangle\psi$)  as a function of charge length($q$) in RN geometry for an arbitrarily selected pulsar PSRB 1937+21.}\label{fig:4}
\end{figure*}

\begin{figure*}[!htb]\centering
\includegraphics{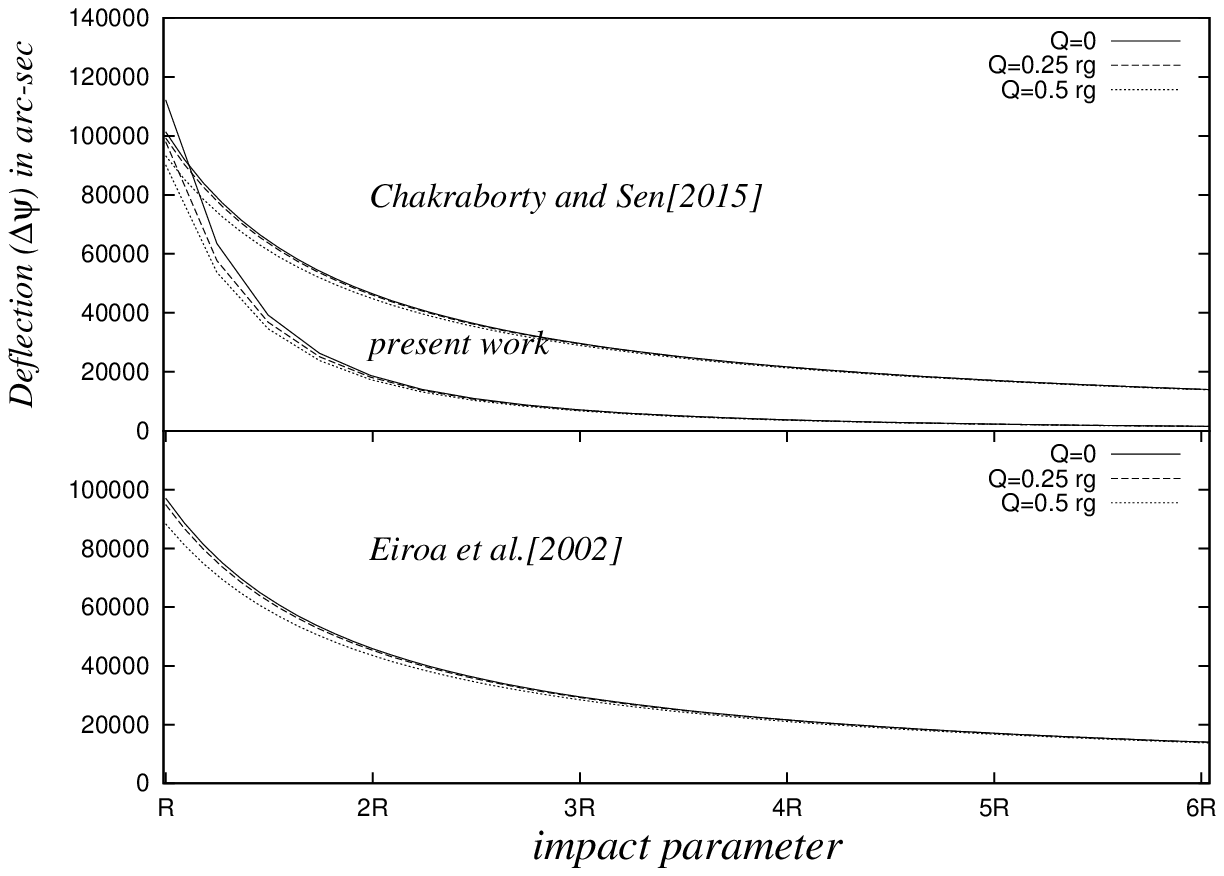}% Here is how to import EPS art 3
\caption{Deflection($\triangle\psi$)  as a function of impact parameter in RN geometry for an arbitrarily selected pulsar PSRB 1937+21.}\label{fig:5}
\end{figure*}

\begin{figure*}[!htb]\centering
\includegraphics{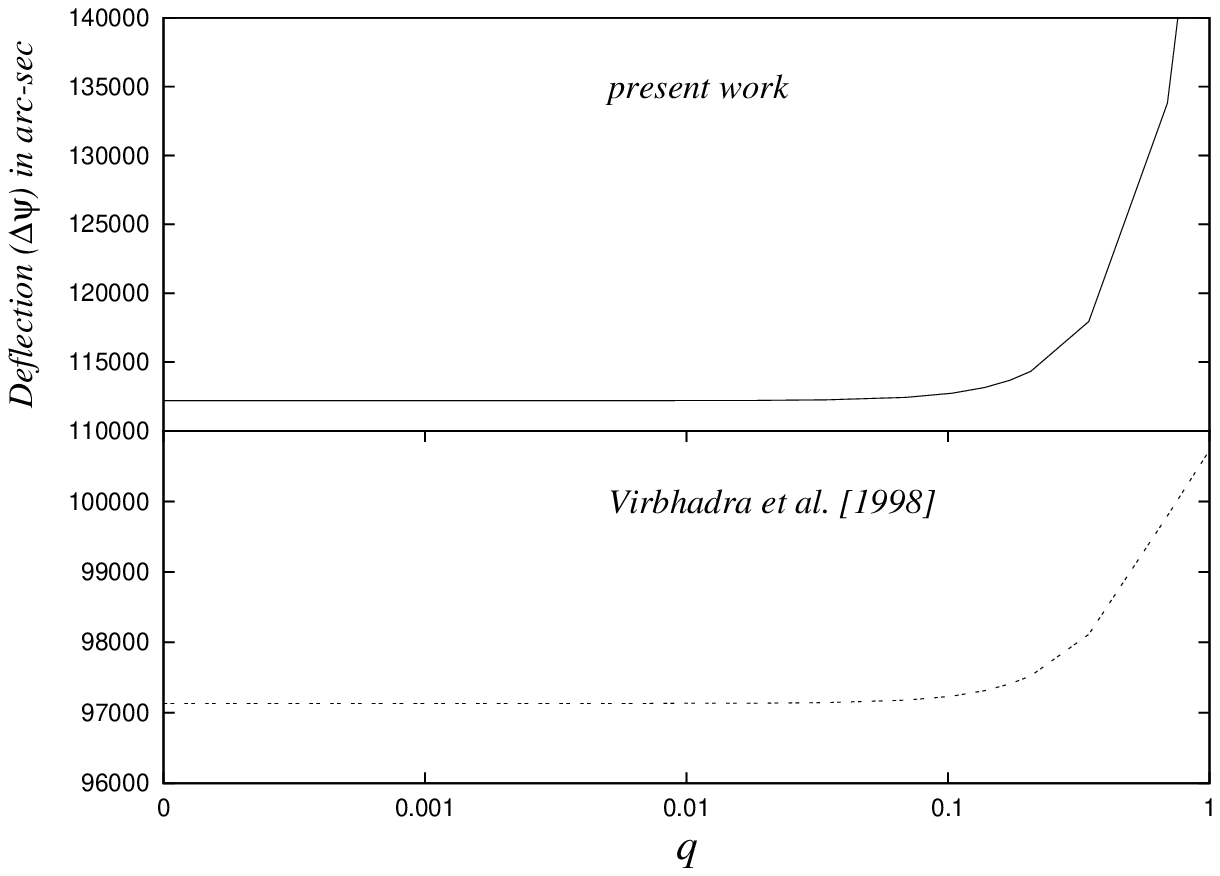}% Here is how to import EPS art 3
\caption{Deflection($\triangle\psi$)    as a function of $q$ in JNW geometry for an arbitrarily selected pulsar PSRB 1937+21.}\label{fig:6}
\end{figure*}

\begin{figure*}[!htb]\centering
\includegraphics{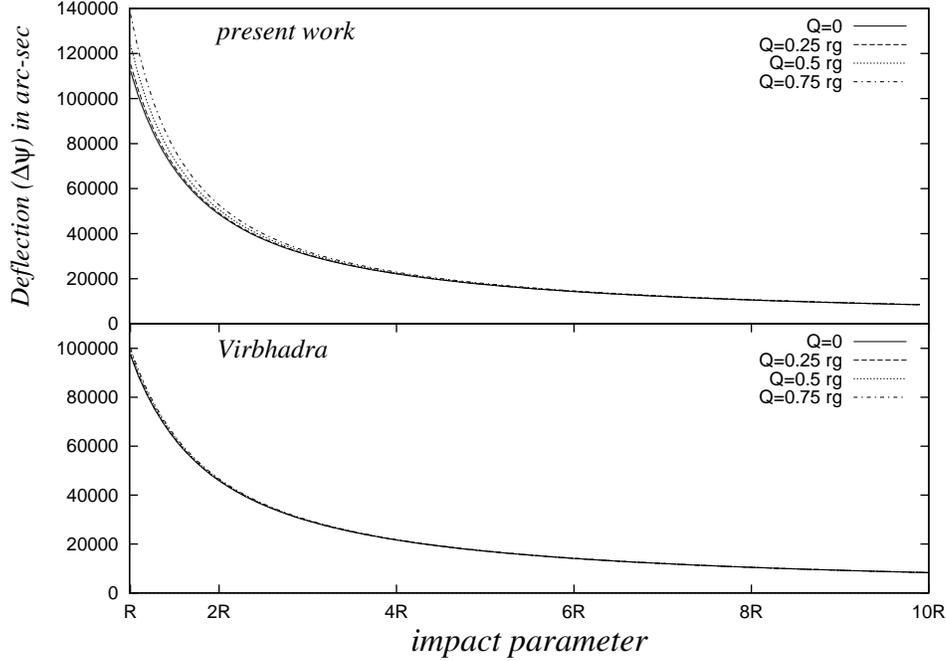}% Here is how to import EPS art 3
\caption{Deflection($\triangle\psi$)   as a function of impact parameter in JNW geometry for an arbitrarily selected pulsar PSRB 1937+21.}\label{fig:7}
\end{figure*}

\subsubsection{\label{2.4.1} RN geometry : }

Sereno in 2004\cite{jr34}
studied the gravitational lensing by RN black hole in the weak field limit
in quasi-Minkovskian co-ordinate. The author followed the Fermat's principle
and obtained the approximated value of refractive index in
quasi-Minkovskian co-ordinate(eqn.8\cite{jr34}). But here we have calculated
the exact value of refractive index in Boyer and Lindquist co-ordinate due to RN
black hole. In 2002 Eiroa {\it et al.}\cite{jr11} also studied the strong gravitational lensing
by RN black hole using null geodesic method in Boyer-Lindquist co-ordinate upto 2nd order of charge(eqn.55\cite{jr11}).
Very recently S. Chakraborty and A. K. Sen\cite{jr41} also
studied the deflection of light ray by Kerr-Newman geometry, using null geodesic method
in Boyer-Lindquist co-ordinate upto 4th order of charge and mass.
If we put rotation parameter equal to zero, we get the the defection angle due to RN geometry(eqn.37\cite{jr41}).
In fig.\ref{fig:4} and fig.\ref{fig:5} we have plotted the deflection angle ($\triangle\psi$) as a function of charge radius ($q$) and impact parameter respectively using the expression of different authors.
In both the figures the set of curves follow similar pattern, but the difference
lies in the magnitude of deflection values. The reason  could be due to the fact that,
Eiroa {\it et al.}\cite{jr11} calculated upto 2nd order of charge and
Chakraborty and Sen\cite{jr41} calculated upto 4th order of charge and mass. But in the present work
we have not used any approximation to calculate the deflection angle. So our calculated values are
claimed to be most exact so far.

\subsubsection{\label{2.4.2} JNW geometry : }

In 1998 Virbhadra {\it et al.}\cite{jr10} calculated the Einstein deflection angle (up to second order)
(eqn.24\cite{jr10})with JNW space-time using null geodesic method.
In Fig.\ref{fig:6} and Fig.\ref{fig:7} we have plotted deflection angle as a function of charge radius
and impact parameter calculated by Virbhadra {\it et al.}\cite{jr10} and present work.
Here also all the curves follow similar pattern, but the values of deflection angle differ between the curves.
Virbhadra {\it et al.}\cite{jr10} calculated only up to second order and we have not considered any approximation
in our present calculation. So here again, our calculated values are claimed to be most exact.

From the above discussions, we may notice that the RN space-time and JNW space-time
both are not similar kind of space-time although both represents the static solution of
Einstein-Maxwell field equation for a charged, non-rotating and spherically symmetric
gravitating body. With the increase of scalar charge the deflection angle decreases
due to RN geometry and increases due to JNW geometry.

\section{{\label{3}}Discussion and conclusions}

In this paper we have presented the
light deflection angle  due to a charged gravitating body in
Reissner-Nordstr\"{o}m space time and Janis, Newman and Winicour
space time, following \textit{Material Medium Approach}.
We have calculated the refractive index and deflection angle
for both the space time,
without using any weak field approximation. The plots of deflection angle against
charge and impact parameter, show the same pattern as obtained by the others using most
conventional method of \textit{Null geodesic}.
From the results obtained in the present work, it is concluded that the bending angle
decreases with the increase of charge in RN space time and increases with
charge in JNW space time.

\begin{acknowledgments}
SR acknowledge Trishna Bordaloi and  Samujwal Das, M.Sc. Student of Assam University, Silchar
who had done M.Sc. project under this topic. SR deeply acknowledge S Chakraborty, Assam University,
Silchar also for useful discussions. Finally we acknowledge
grants from UGC-SAP under which the work was done.
\end{acknowledgments}

\appendix

\section{\label{A}Bending angle due to RN geometry}

We rewrite the Eqn.(\ref{E22}) as

\begin{equation}\label{A1}
    \triangle \psi = 2I - \pi
\end{equation}

where

\begin{eqnarray}\label{A2}
I&=&n(v,q) v \int^{\infty}_{v}  { \frac {dx}{x
\sqrt{(n(x,q) x)^2-(n(v,q) v)^2}}}\nonumber\\
&=&D \int^{\infty}_{v}  { \frac {dx}{x
\sqrt{(n(x,q) x)^2- D^2}}}
\end{eqnarray}

with $D = n(v,q) v$.

To evaluate the above integral, we follow a procedure similar to what
was done by Sen\cite{jr36} and Roy and Sen\cite{jr37}.

Thus with the value of refractive index from Eqn.(\ref{E17}) and $D = D_r $
(where $D_r = n(v,q) v$, $ n(v,q)$ is the refractive index due to RN geometry
at the limit of impact parameter) we have

\begin{widetext}
\begin{eqnarray}\label{A3}
I=&& D_{r} \int^{\infty}_{v} { \frac {dx}{x\sqrt{\{n_{0}(x)  (1+ C_{x})^{-1}\cdot x\}^2- D_{r}^2}}}\nonumber\\
=&& D_{r} \int^{\infty}_{v} \frac{ dx}{x\sqrt{n_{0}^{2}(x) x^{2} -D_{0}^2+n_{0}^{2}(x) x^{2} (1+ C_{x})^{-2}- n_{0}^{2}(x) x^{2}+D_{0}^2-D_{r}^2  }}\nonumber\\
=&& D_{r} \int^{\infty}_{v} \frac{ dx}{x\sqrt{n_{0}^{2}(x) x^{2}-D_{0}^2}}[1 + \frac{n_{0}^{2}(x) x^{2} \{(1+ C_{x})^{-2}  -  1\} +D_{0}^2-D_{r}^2 }{n_{0}^{2}(x) x^{2}-D_{0}^2}]^{-\frac{1}{2}}\nonumber\\
=&& D_{r} \int^{\infty}_{v} \frac{ dx}{x\sqrt{n_{0}^{2}(x) x^{2}-D_{0}^2}}[1 + K(x)]^{-\frac{1}{2}}
\end{eqnarray}
\end{widetext}

where  $D_{0}=n_{0}(v).v$ (corresponding to schwarzschild
deflection). And we have also denoted

\begin{eqnarray}\label{A4}
K(x)= \frac{n_{0}^{2}(x) x^{2} \{(1+ C_{x})^{-2}  -  1\} +D_{0}^2-D_{r}^2 }{n_{0}^{2}(x) x^{2}-D_{0}^2}
\end{eqnarray}

Here, we can show that $ K(x)<< 1$.
To evaluate the value of $K(x)$ of Eqn.(\ref{A4}), the
value of $((1+ C_{x})^{-2}  -  1)$ is as follows:

\begin{eqnarray}\label{A5}
  (1+ C_{x})^{-2}  -  1 =&& \frac{1}{(1+ C_{x})^2} -1 \nonumber\\
  =&& \frac{1}{(1+ \frac{q^2}{x(x-1)})^2} -1 \nonumber\\
  =&& \frac{-\frac{2q^2x(x-1)+q^4}{x^2(x-1)^2}}{\frac{(x(x-1)+q^2)^2}{x^2(x-1)^2}}\nonumber\\
  =&& -\frac{2q^2x(x-1)+q^4}{(x(x-1)+q^2)^2}
\end{eqnarray}

Substituting the value of $((1+ C_{x})^{-2}  -  1)$ from Eqn.(\ref{A5})
and $n_0(x)=x/(x-1)$ we can write the value of $K(x)$ as

\begin{eqnarray}\label{A6}
K(x)=&&\frac{n_{0}^{2}(x) x^{2} \{(1+ C_{x})^{-2}  -  1\}+D_{0}^2-D_{r}^2 }{n_{0}^{2}(x) x^{2}-D_{0}^2}\nonumber\\
=&&\frac{n_{0}^{2}(x) x^{2} \{(1+ C_{x})^{-2}  -  1\}+D_{0}^2-D_{0}^2 (1+ C_{v})^{-2} }{n_{0}^{2}(x) x^{2}-D_{0}^2}\nonumber\\
=&&\frac{  \frac{x^{4}}{(x-1)^2} \{-\frac{2q^2x(x-1)+q^4}{(x(x-1)+q^2)^2} \}+D_{0}^2\{\frac{2q^2v(v-1)+q^4}{(v(v-1)+q^2)^2}\} }{\frac{x^{4}}{(x-1)^2}-D_{0}^2}\nonumber\\
=&&\frac{ -  x^{4} \{\frac{2q^2x(x-1)+q^4}{(x(x-1)+q^2)^2} \}+D_{0}^2 (x-1)^2\{\frac{2q^2v(v-1)+q^4}{(v(v-1)+q^2)^2}\} }{ x^{4}-D_{0}^2 (x-1)^2}\nonumber\\
\end{eqnarray}

At this stage we can show that $ K(x)<< 1$.
 As $K(x)$ is discontinuous at $x = v$, we can remove its discontinuity and
evaluate its value  by applying L'Hospital's rule.

Therefore, from Eqns.(\ref{A1}) and (\ref{A3}) one can write:

\begin{widetext}
\begin{eqnarray}\label{A7}
\triangle \psi =& 2 D_{r} \int^{\infty}_{v} \frac{ dx}{x\sqrt{n_{0}^{2}(x) x^{2}-D_{0}^2}}[1 -\frac{1}{2} K(x)+\frac{3}{8}K^{2}(x)-\frac{5}{16}K^{3}(x)+\frac{35}{128}K^{4}(x)-\frac{63}{254}K^{5}(x)+........] - \pi\nonumber\\
=& 2[I_0+I_1+I_2+I_3+............]  - \pi
\end{eqnarray}
\end{widetext}

where, we have introduced some other notations:

\begin{subequations}
\begin{equation}\label{A8a}
I_0= D_r \int ^{\infty}_{v}\frac{ dx}{x\sqrt{n_{0}^{2}(x)
x^{2}-D_{0}^2}}
\end{equation}

\begin{equation}\label{A8b}
I_1= D_r \int ^{\infty}_{v}\frac{ dx}{x\sqrt{n_{0}^{2}(x)
x^{2}-D_{0}^2}}(-1/2 K(x))
\end{equation}

\begin{equation}\label{A8c}
I_2= D_r \int ^{\infty}_{v}\frac{ dx}{x\sqrt{n_{0}^{2}(x)
x^{2}-D_{0}^2}}(3/8 K^2(x))
\end{equation}
\end{subequations}

and so on.

Now $I_0$ can be evaluate by following the same procedure as
Sen\cite{jr36} and Roy and Sen\cite{jr37}. According to
Roy and Sen\cite{jr37} $I_0$ can be split into two integrals
as $I_{01}$ and $I_{02}$. Here, $D_k$ is replaced by $D_r$.
Thus the value of $I_{01}$ and $I_{02}$ will be :

\begin{equation}\label{A9}
    I_{01} = \frac{D_{r}}{D_{0}}\cdot \frac{\pi}{2}
\end{equation}
and
\begin{equation}\label{A10}
    I_{02} = D_{r}\int ^{a}_{0}\frac{ z dz}{ \sqrt{1-D_{0}^2z^{2}(1-z)^{2}}}
\end{equation}

where we change the variable as
$z=\frac{1}{x}$, so that the limits of this integration changes
from $z=\frac{1}{v}=a (say)$ to $z=0$. The integral (\ref{A10}) can
be evaluated in terms of Elliptical function
as expressed by Eqn.(18) of Sen\cite{jr36} and finally for
a given value of $a$, its numerical value can be obtained as
Roy and Sen\cite{jr37}.

Now substituting the value of $K(x)$ from Eqn.(\ref{A6}),
$n_0(x)=\frac{x}{x-1}$ and applying the change of variable
as $z=\frac{1}{x}$ the integral $I_1$  becomes

\begin{widetext}
\begin{eqnarray}\label{A11}
I_{1}=&&- \frac{1}{2} D_{r}\int^{\infty}_{v} \frac{ K(x)}{x\sqrt{n_{0}^{2} x^{2}-D_{0}^2}}dx\nonumber\\
=&& - \frac{1}{2} D_{r}\int^{\infty}_{v} \frac{ 1}{x\sqrt{  \frac{x^{4}}{(x-1)^{2}}-D_{0}^2}}
[\frac{ -  x^{4} \{\frac{2q^2x(x-1)+q^4}{(x(x-1)+q^2)^2} \}+D_{0}^2 (x-1)^2\{\frac{2q^2v(v-1)+q^4}{(v(v-1)+q^2)^2}\} }{ x^{4}-D_{0}^2 (x-1)^2}]dx\nonumber\\
=&&-  \frac{1}{2} D_{r}\int^{\infty}_{v} \frac{(x- 1)}{\sqrt{   x^{6}-D_{0}^2 x^{2}(x-1)^{2}}}
[\frac{ -  x^{4} \{\frac{2q^2x(x-1)+q^4}{(x(x-1)+q^2)^2} \}+D_{0}^2 (x-1)^2\{\frac{2q^2v(v-1)+q^4}{(v(v-1)+q^2)^2}\} }{ x^{4}-D_{0}^2 (x-1)^2}]dx \nonumber\\
=&&-  \frac{1}{2}D_{r}\int^{a}_{0} \frac{( 1 -z)}{\sqrt{   1 -D_{0}^2 z^{2}(1 -z)^{2}}}
[\frac{ -   \frac{2q^2z^2 (1-z)+q^4z^4}{((1-z)+q^2z^2)^2}  +D_{0}^2 z^2(1-z)^2\{\frac{2q^2v(v-1)+q^4}{(v(v-1)+q^2)^2}\} }{ 1-D_{0}^2 z^2(1-z)^2}]dz\nonumber\\
\end{eqnarray}

Applying the same procedure, the other integrals $I_2$, $I_3$ etc. are as follows:

\begin{eqnarray}\label{A12}
    I_{2} =  \frac{3}{8}D_{r}\int^{a}_{0} \frac{( 1 -z)}{\sqrt{   1 -D_{0}^2 z^{2}(1 -z)^{2}}}
[\frac{ -   \frac{2q^2z^2 (1-z)+q^4z^4}{((1-z)+q^2z^2)^2}  +D_{0}^2 z^2(1-z)^2\{\frac{2q^2v(v-1)+q^4}{(v(v-1)+q^2)^2}\} }{ 1-D_{0}^2 z^2(1-z)^2}]^2 dz\nonumber\\
\end{eqnarray}

\begin{eqnarray}\label{A13}
    I_{3} =   -  \frac{5}{16}D_{r}\int^{a}_{0} \frac{( 1 -z)}{\sqrt{   1 -D_{0}^2 z^{2}(1 -z)^{2}}}
[\frac{ -   \frac{2q^2z^2 (1-z)+q^4z^4}{((1-z)+q^2z^2)^2}  +D_{0}^2 z^2(1-z)^2\{\frac{2q^2v(v-1)+q^4}{(v(v-1)+q^2)^2}\} }{ 1-D_{0}^2 z^2(1-z)^2}]^3 dz\nonumber\\
\end{eqnarray}

\begin{eqnarray}\label{A14}
    I_{4} =  \frac{35}{64}D_{r}\int^{a}_{0} \frac{( 1 -z)}{\sqrt{   1 -D_{0}^2 z^{2}(1 -z)^{2}}}
[\frac{ -   \frac{2q^2z^2 (1-z)+q^4z^4}{((1-z)+q^2z^2)^2}  +D_{0}^2 z^2(1-z)^2\{\frac{2q^2v(v-1)+q^4}{(v(v-1)+q^2)^2}\} }{ 1-D_{0}^2 z^2(1-z)^2}]^4 dz\nonumber\\
\end{eqnarray}

\begin{eqnarray}\label{A15}
    I_{5} =   -  \frac{63}{256}D_{r}\int^{a}_{0} \frac{( 1 -z)}{\sqrt{   1 -D_{0}^2 z^{2}(1 -z)^{2}}}
[\frac{ -   \frac{2q^2z^2 (1-z)+q^4z^4}{((1-z)+q^2z^2)^2}  +D_{0}^2 z^2(1-z)^2\{\frac{2q^2v(v-1)+q^4}{(v(v-1)+q^2)^2}\} }{ 1-D_{0}^2 z^2(1-z)^2}]^5 dz\nonumber\\
\end{eqnarray}

Therefore, substituting all the values of $I_0$, $I_1$, $I_2$, $I_3$ etc.
the Eqn.(\ref{A7}) becomes

\begin{eqnarray}\label{A16}
  \triangle \psi =&& 2  [\frac{D_{r}}{D_{0}} \frac{\pi}{2}  + D_{r}\{\int ^{a}_{0}\frac{ z dz}{ \sqrt{1-D_{0}^2z^{2}(1-z)^{2}}}\nonumber\\
&&-  \frac{1}{2} \int^{a}_{0} \frac{( 1 -z)}{\sqrt{   1 -D_{0}^2 z^{2}(1 -z)^{2}}}
[\frac{ -   \frac{2q^2z^2 (1-z)+q^4z^4}{((1-z)+q^2z^2)^2}  +D_{0}^2 z^2(1-z)^2\{\frac{2q^2v(v-1)+q^4}{(v(v-1)+q^2)^2}\} }{ 1-D_{0}^2 z^2(1-z)^2}]dz\nonumber\\
&&+ \frac{3}{8} \int^{a}_{0} \frac{( 1 -z)}{\sqrt{   1 -D_{0}^2 z^{2}(1 -z)^{2}}}
[\frac{ -   \frac{2q^2z^2 (1-z)+q^4z^4}{((1-z)+q^2z^2)^2}  +D_{0}^2 z^2(1-z)^2\{\frac{2q^2v(v-1)+q^4}{(v(v-1)+q^2)^2}\} }{ 1-D_{0}^2 z^2(1-z)^2}]^2 dz\nonumber\\
&& -  \frac{5}{16} \int^{a}_{0} \frac{( 1 -z)}{\sqrt{   1 -D_{0}^2 z^{2}(1 -z)^{2}}}
[\frac{ -   \frac{2q^2z^2 (1-z)+q^4z^4}{((1-z)+q^2z^2)^2}  +D_{0}^2 z^2(1-z)^2\{\frac{2q^2v(v-1)+q^4}{(v(v-1)+q^2)^2}\} }{ 1-D_{0}^2 z^2(1-z)^2}]^3 dz\nonumber\\
&& +  \frac{35}{64} \int^{a}_{0} \frac{( 1 -z)}{\sqrt{   1 -D_{0}^2 z^{2}(1 -z)^{2}}}
[\frac{ -   \frac{2q^2z^2 (1-z)+q^4z^4}{((1-z)+q^2z^2)^2}  +D_{0}^2 z^2(1-z)^2\{\frac{2q^2v(v-1)+q^4}{(v(v-1)+q^2)^2}\} }{ 1-D_{0}^2 z^2(1-z)^2}]^4 dz\nonumber\\
&& -  \frac{63}{256} \int^{a}_{0} \frac{( 1 -z)}{\sqrt{   1 -D_{0}^2 z^{2}(1 -z)^{2}}}
[\frac{ -   \frac{2q^2z^2 (1-z)+q^4z^4}{((1-z)+q^2z^2)^2}  +D_{0}^2 z^2(1-z)^2\{\frac{2q^2v(v-1)+q^4}{(v(v-1)+q^2)^2}\} }{ 1-D_{0}^2 z^2(1-z)^2}]^5 dz\nonumber\\
&&+ ........\}]-\pi\nonumber\\
=&& (\frac{D_{r}}{D_{0}}-1) \pi   + 2D_{r}\{\int ^{a}_{0}\frac{ z dz}{ \sqrt{1-D_{0}^2z^{2}(1-z)^{2}}}\nonumber\\
&&-  \frac{1}{2} \int^{a}_{0} \frac{( 1 -z)}{\sqrt{   1 -D_{0}^2 z^{2}(1 -z)^{2}}}
[\frac{ -   \frac{2q^2z^2 (1-z)+q^4z^4}{((1-z)+q^2z^2)^2}  +D_{0}^2 z^2(1-z)^2\{\frac{2q^2v(v-1)+q^4}{(v(v-1)+q^2)^2}\} }{ 1-D_{0}^2 z^2(1-z)^2}]dz\nonumber\\
&& + \frac{3}{8} \int^{a}_{0} \frac{( 1 -z)}{\sqrt{   1 -D_{0}^2 z^{2}(1 -z)^{2}}}
[\frac{ -   \frac{2q^2z^2 (1-z)+q^4z^4}{((1-z)+q^2z^2)^2}  +D_{0}^2 z^2(1-z)^2\{\frac{2q^2v(v-1)+q^4}{(v(v-1)+q^2)^2}\} }{ 1-D_{0}^2 z^2(1-z)^2}]^2 dz\nonumber\\
&& -  \frac{5}{16} \int^{a}_{0} \frac{( 1 -z)}{\sqrt{   1 -D_{0}^2 z^{2}(1 -z)^{2}}}
[\frac{ -   \frac{2q^2z^2 (1-z)+q^4z^4}{((1-z)+q^2z^2)^2}  +D_{0}^2 z^2(1-z)^2\{\frac{2q^2v(v-1)+q^4}{(v(v-1)+q^2)^2}\} }{ 1-D_{0}^2 z^2(1-z)^2}]^3 dz\nonumber\\
&& +  \frac{35}{64} \int^{a}_{0} \frac{( 1 -z)}{\sqrt{   1 -D_{0}^2 z^{2}(1 -z)^{2}}}
[\frac{ -   \frac{2q^2z^2 (1-z)+q^4z^4}{((1-z)+q^2z^2)^2}  +D_{0}^2 z^2(1-z)^2\{\frac{2q^2v(v-1)+q^4}{(v(v-1)+q^2)^2}\} }{ 1-D_{0}^2 z^2(1-z)^2}]^4 dz\nonumber\\
&& -  \frac{63}{256} \int^{a}_{0} \frac{( 1 -z)}{\sqrt{   1 -D_{0}^2 z^{2}(1 -z)^{2}}}
[\frac{ -   \frac{2q^2z^2 (1-z)+q^4z^4}{((1-z)+q^2z^2)^2}  +D_{0}^2 z^2(1-z)^2\{\frac{2q^2v(v-1)+q^4}{(v(v-1)+q^2)^2}\} }{ 1-D_{0}^2 z^2(1-z)^2}]^5 dz\nonumber\\
&&+ ........\}
\end{eqnarray}
\end{widetext}

The above expression represents the light deflection angle
due to charged gravitating mass in RN space time.

\section{\label{B} Bending angle due to JNW geometry}

Here also we will follow the same procedure as Appendix \ref{A}.
Now with the value of refractive index from Eqn.(\ref{E18}) and $D = D_j $
(where $D_j = n(v,q) v $, $ n(v,q)$ is the refractive index due to JNW geometry
at the limit of impact parameter), we have
\begin{widetext}
\begin{eqnarray}\label{B1}
I_j =&& D_j \int^{\infty}_{v} \frac{dx}{x \sqrt{(n(x,q)x )^2 - D_j^2 }}\nonumber\\
=&& D_j \int^{\infty}_{v} \frac{dx}{x \sqrt{(\frac{1}{1-\frac{1}{x}\sqrt{1+4q^2}})^{2/\sqrt{1+4q^2}} x^2 - D_j^2}}\nonumber\\
=&& D_j \int^{\infty}_{v} \frac{(1- \frac{1}{x}\sqrt{1+4q^2})^{1/  \sqrt{1+4q^2}}}{x^2 \sqrt{1 - \frac{D_j^2}{x^2} (1-\frac{1}{x}\sqrt{1+4q^2})^{2/\sqrt{1+4q^2}}}} dx \nonumber\\
=&& D_j \int^{\infty}_{v} \frac{  (1- \frac{1}{x}\sqrt{1+4q^2})^{1/\sqrt{1+4q^2}} - \frac{1}{x} (1- \frac{1}{x}\sqrt{1+4q^2})^{\frac{1}{ \sqrt{1+4q^2}}-1}}{x^2\sqrt{1 - \frac{D_j^2}{x^2} (1-\frac{1}{x}\sqrt{1+4q^2})^{2/\sqrt{1+4q^2}}}} dx \nonumber\\
&& +  D_j \int^{\infty}_{v} \frac{  \frac{1}{x} (1- \frac{1}{x}\sqrt{1+4q^2})^{\frac{1}{ \sqrt{1+4q^2}}-1} }{x^2\sqrt{1 - \frac{D_j^2}{x^2} (1-\frac{1}{x}\sqrt{1+4q^2})^{2/\sqrt{1+4q^2}}}} dx\nonumber\\
=&& I_{j1} + I_{j2}
\end{eqnarray}

Now, let

\begin{equation*}
    y= \frac{D_j}{x} (1- \frac{1}{x} \sqrt{1+4q^2})^{1/\sqrt{1+4q^2}}
\end{equation*}

 so that

 \begin{eqnarray*}
 dy &=& - \frac{D_j}{x^2}  (1- \frac{1}{x}\sqrt{1+4q^2})^{1/\sqrt{1+4q^2}} dx +
  \frac{D_j}{x^3}  (1- \frac{1}{x}\sqrt{1+4q^2})^{\frac{1}{\sqrt{1+4q^2}}-1}dx\\
   &=& - \frac{ D_j}{x^2} [ (1- \frac{1}{x}\sqrt{1+4q^2})^{1/\sqrt{1+4q^2}} -
   \frac{1}{x} (1- \frac{1}{x}\sqrt{1+4q^2})^{\frac{1}{\sqrt{1+4q^2}}-1}]dx\\
 \end{eqnarray*}

Thus the limit changes to $y=0$ and $y= \frac{D_j}{v}(1- \frac{1}{v}\sqrt{1+4q^2})^{1/\sqrt{1+4q^2}} = D_j \frac{1}{D_j} = 1$ as x changes to $v$ and $\infty$.
So,

\begin{eqnarray}\label{B2}
I_{j1}&=& D_j \int^{\infty}_{v} \frac{  (1- \frac{1}{x}\sqrt{1+4q^2})^{1/\sqrt{1+4q^2}} - \frac{1}{x} (1- \frac{1}{x}\sqrt{1+4q^2})^{\frac{1}{ \sqrt{1+4q^2}}-1}}{x^2\sqrt{1 - \frac{D_j^2}{x^2} (1-\frac{1}{x}\sqrt{1+4q^2})^{2/\sqrt{1+4q^2}}}} dx \nonumber\\
&=& \int^{1}_{0} \frac{dy}{\sqrt{1- y^2}}\nonumber\\
&=& \frac{\pi}{2}
\end{eqnarray}
\end{widetext}
Now by applying the change of variable as $z=\frac{1}{x}$ like Appendix \ref{A} we may
write the above integral $I_{j2}$ as

\begin{eqnarray}\label{B3}
  I_{j2} &=& D_j \int^{\infty}_{v} \frac{  \frac{1}{x} (1- \frac{1}{x}\sqrt{1+4q^2})^{\frac{1}{ \sqrt{1+4q^2}}-1} }{x^2\sqrt{1 - \frac{D_j^2}{x^2} (1-\frac{1}{x}\sqrt{1+4q^2})^{2/\sqrt{1+4q^2}}}} dx\nonumber \\
   &=&  D_j \int^{a}_{0} \frac{ z (1- z \sqrt{1+4q^2})^{\frac{1}{ \sqrt{1+4q^2}}-1} }{ \sqrt{1 -  D_j^2 z^2 (1- z \sqrt{1+4q^2})^{2/\sqrt{1+4q^2}}}} dz\nonumber\\
\end{eqnarray}

Thus from expression (\ref{A1}) and (\ref{B1}) the light deflection angle due to JNW space time
can be written as

\begin{eqnarray}\label{B4}
  \triangle \psi &=& 2[\frac{\pi}{2} +  D_j \int^{a}_{0} \frac{ z (1- z \sqrt{1+4q^2})^{\frac{1}{ \sqrt{1+4q^2}}-1} }{ \sqrt{1 -  D_j^2 z^2 (1- z \sqrt{1+4q^2})^{2/\sqrt{1+4q^2}}}} dz] - \pi \nonumber\\
   &=& 2 D_j \int^{a}_{0} \frac{ z (1- z \sqrt{1+4q^2})^{\frac{1}{ \sqrt{1+4q^2}}-1} }{ \sqrt{1 -  D_j^2 z^2 (1- z \sqrt{1+4q^2})^{2/\sqrt{1+4q^2}}}} dz
\end{eqnarray}

\nocite{*}
%\newpage %Just because of unusual number of tables stacked at end
%\bibliography{apssamp}
%\bibliography{basename of .bib file}% Produces the bibliography via BibTeX.

\bf{References}

%\end{references}

\end{document}